\documentstyle[11pt,moriond,epsfig]{article}

\def\Journal#1#2#3#4{{#1} {\bf #2}, #3 (#4)}

\def\NPB{{\em Nucl. Phys.} B}
\def\PLB{{\em Phys. Lett.}  B}
\def\PRL{\em Phys. Rev. Lett.}
\def\PRD{{\em Phys. Rev.} D}

\def\PhysRep{{\em Phys. Rept.}}
\def\EPJC{{\em Eur. Phys. J.} C}
\def\ProgTh{\em Prog. Theor. Phys.}
\def\RMP{\em Rev. Mod. Phys.}
\def\NPProc{\em Nucl. Phys. Proc. Suppl.}
\def\JETPLet{\em JETP Lett.}
\def\JETP{\em J. Exp. Theor. Phys. }
\def\JHEP{\em JHEP}
\def\IJMP{{\em Int. J. Mod. Phys.} A}
\def\parsym{\stackrel{\leftrightarrow}{\partial}}
\def\ra{\rightarrow}

\def\omg{\omega}

\def\be{\begin{equation}}
\def\ee{\end{equation}}
\def\bea{\begin{eqnarray}}
\def\eea{\end{eqnarray}}

\def\bb{\bibitem}
\begin{document}
\vspace*{4cm}
\title{EFFECTS OF THE ($\rho$, $\omg$, $\phi$) MIXING ON THE DIPION MASS SPECTRUM
IN $e^+e^-$ ANNIHILATION AND $\tau$ DECAY}

\author{ M. BENAYOUN }

\address{LPNHE Paris VI/VII, IN2P3/CNRS, F-75252 Paris, France
\\[0.5cm] {\rm in collaboration with P.~DAVID, L.~ DELBUONO, O.~LEITNER and H.~B.~O'CONNELL} }

\maketitle\abstracts{
A way to explain the
puzzling difference between the pion form factor as measured in $e^+e^-$ annihilations
and in $\tau$ decays is discussed. We show that isospin symmetry breaking, beside the already
identified effects, produces also a full mixing  between the  $\rho^0$, $\omg$ and $\phi$
mesons which generates an isospin 0 component inside the $\rho^0$ meson. This effect, not accounted
for in current treatments of the problem, seems able to account for the apparent mismatch
between $e^+e^-$ and $\tau$ data below the $\phi$ mass.
}

\section{Introduction}
\label{sect1}
\indent \indent In order to get  a theoretical estimate of the muon anomalous
magnetic moment $g-2$, one  needs to estimate precisely the photon vacuum 
polarization  (see Jegerlehner~\cite{Fred} for a comprehensive review). Its leptonic part 
can be computed theoretically to a high precision from QED, but the dominance
of non--pertubative effects in the low energy region prevents 
to perform likewise starting from QCD in order to estimate the hadronic part.  This is 
instead done by means of a dispersion integral involving the measured cross section 
$\sigma(e^+ e^-\rightarrow {\rm hadrons})$; however, the integration kernel 
is such that the low energy region contribution is enhanced by a $\sim 1/s^2$ factor. Because 
of this, the non--pertubative region provides, by far, the largest contribution
to the hadronic vacuum polarization (VP). Additionally, the annihilation process
$e^+ e^-\rightarrow \pi^+ \pi^-$ alone happens to provide more than 60 \% of the
total hadronic VP.

As one has to rely on data in order to estimate  this contribution, 
the precision
of the measured $\sigma(e^+ e^-\rightarrow \pi^+ \pi^-)$ is clearly 
an important issue. For this purpose,
several sets of data collected in $e^+ e^-$
annihilations at low energies during a long period of time are available. This covers
the former data sets  collected  by the OLYA, CMD and DM1 Collaborations -- which are gathered
in the review by  Barkov {\it et al}~\cite{Barkov} --, and the data
sets  more recently collected by the CMD2~\cite{CMD2-1995corr,CMD2-1998-1,CMD2-1998-2} 
and SND~\cite{SND} Collaborations. Additional data sets taking advantage
of the initial state radiation mechanism have also been collected by the
KLOE, BaBar and Belle Collaborations and are expected to become available soon.

Moreover, high statistics data on the decay $\tau^\pm \ra \nu_\tau \pi^\pm \pi^0$
are also available from the ALEPH~\cite{Aleph} and CLEO~\cite{Cleo} 
Collaborations. As the pion form factor ({\it i.e.} the dipion mass spectrum) 
in $\tau$ decays and in $e^+ e^-$ annilhilations are related by the Conserved
Vector Current (CVC) assumption, these data are expected to be useful in order
to improve the estimate of the photon hadronic VP. Indeed, 
these two kinds of data can only differ by isospin breaking effects 
which are subject to accurate estimates.

Isospin symmetry breaking effects have been especially studied in order 
to include the $\tau$ data in the estimation of the photon hadronic 
VP. This covers non trivial effects specific of the $\tau$ decay like the short
range~\cite{Marciano} and long range~\cite{Cirigliano,Mexico} isospin breaking 
factors, but also more standard effects easier to account for~: mass differences between 
charged and neutral 
pions and charged and neutral $\rho$ mesons, the $\rho^\pm-\rho^0$ width difference, 
or the $\omg$ and $\phi$ contributions 
to the  $e^+ e^-$ annihilation amplitude (see, for instance, the work by 
Davier {\it et al.}~\cite{DavierPrevious,DavierPrevious3}). 

As a preliminary step in the process of
including $\tau$ data in estimating the photon hadronic VP, the comparision
has been done of the pion form factor as measured in  $e^+ e^-$ annihilations
and as derived from $\tau$ decays while accounting for all known isospin breaking effects
appropriately~\cite{DavierPrevious,Fred,DavierPrevious3}. This 
comparison, however, clearly exhibits an unexpected $s$--dependence of the difference between
the  $e^+ e^-$ data and the pion form factor function reconstructed from
$\tau$ data, as reported still recently by Davier~\cite{Davier2007}~\footnote{See also A. 
Hoecker, {\sf http://moriond.in2p3.fr/QCD/2008/MorQCD08Prog.html}, 
Moriond QCD, March 2008.\label{Hoecker1}}. 

This  mismatch is an important issue as the photon hadronic VP as reconstructed
from $e^+ e^-$ data leads to a theoretical prediction for the muon $g-2$
at $\simeq 3.3 ~\sigma$  from its measured value~\cite{BNL}; in contrast, the prediction
derived from $\tau$ data fed with all currently identified isospin symmetry breaking effects
provides an expectation in close agreement~\cite{Davier2007}$^{~\ref{Hoecker1}}$
with the $g-2$ value directly measured
at BNL~\cite{BNL}. Therefore, the question is whether the $(e^+e^--\tau)$ mismatch can be explained
by physics only connected with isospin symmetry breaking -- and then something is missing --
or if it calls for another kind of physics effect (actually hard to identify).  Responding
this question by leaving  $e^+e^-$ data beyond any doubt may point towards a new physics
effect exhibited by the muon anomalous magnetic moment.
 
\section{A missing piece of isospin symmetry breaking~?}
\label{sect2}
\indent \indent As it is clear that all identified (and listed above)
effects produced by isospin symmetry breaking should be considered, the question
is about a possible missing piece in the isospin symmetry breaking procedure
(or a piece not appropriately accounted for). 

Actually, a clue towards the solution  we propose has been given by Maltman~\cite{Maltman}; 
using sum rules derived from an OPE input, he concluded that the $\rho$ part of 
the $e^+ e^-$ form factor data was inconsistent with being isospin 1, in contrast with 
the corresponding information provided by $\tau$ data. This statement implies that either
the quality of the available $e^+ e^-$ data~\footnote{One may note that these data
has now been confirmed by several new and precise measurements.} 
can be questioned or that the $\rho^0$ meson
is not a (pure) isospin 1 object. 

%Let us now very briefly outline the way we used in order
%to explore this second possibility. 

Up to now, the model amplitudes used to describe the neutral and charged $\rho$
mesons may differ by feeding their propagators with different masses and widths
to be fit with data; of course, the pion mass difference is also fed in, together
with the $\omg$ (and $\phi$) meson(s) propagators, generally Breit--Wigner
formulae. However, as the $e^+e^-$ data
clearly exhibit~\cite{Barkov,CMD2-1995corr,CMD2-1998-1,CMD2-1998-2}
the narrow (isospin 1 part of the) $\omg$  interfering with the
broad $\rho^0$, one may ask oneself about the existence of a (broad)
isospin 0 part of the $\rho^0$ meson which might make it differing from its
charged partner beyond genuine mass effects. Stated otherwise, the question
is whether mass and width differences for the $\rho$ mesons exhaust isospin symmetry 
breaking in the pion form factor.

\section{The Pion Form Factor at One Loop}
\label{sect3}
\indent \indent In order to make our statements explicit, we have found it appropriate
to work in the framework of the Hidden Local Symmetry (HLS) 
model~\cite{HLS,HLS2}~\footnote{In this paper, we only outline the method and refer 
the reader to~\cite{newpaper} for detailed information.}. However, our arguments
are certainly valid in most other  VMD--like models. In the HLS model, the pion form factor
for both $e^+ e^-$ annihilation and $\tau$ decay writes~:
\be
\begin{array}{ll}
F_\pi(s) = \displaystyle (1-\frac{a}{2}) - 
\frac{f_\rho g_{\rho \pi \pi}}{D_V(s)},~~~~(
f_\rho=a g f_\pi^2~~~, ~~ \displaystyle g_{\rho \pi \pi} =\frac{a g}{2})~~~,
\end{array}
\label{eq1}
\ee
where $a$ is a parameter specific of this model -- close to 2 -- and $g$
is the universal vector coupling -- close to 5.5 (from QCD
sum rules). $D_V(s)=s-m_\rho^2$ is the inverse $\rho$ bare propagator 
($m_\rho^2=a g^2f_\pi^2$). While including one loop
effects, $D_V(s)$ acquires a pion (and kaon) loop term $\Pi_{\rho}(s)$
which shifts the $\rho$ 
pole off the real $s$--axis. The transition amplitude from $\gamma/W$
to (neutral/charged) $\rho$ is also dressed by loop effects; this 
turns out to perform the change~:
\be
 f_\rho \longrightarrow F_\rho = f_\rho -\Pi_{W/\gamma}(s)
\label{eq2}
\ee
in the expression~\footnote{$\Pi_{\gamma}(s)$ refers
to the pion form factor in $e^+e^-$ annihilation, which will be denoted
$F_\pi^e(s)$, while $\Pi_{W}(s)$ refers to the pion form factor in $\tau$ decay 
correspondingly denoted $F_\pi^\tau(s)$.}  for $F_\pi(s)$.
The 3 loop functions $\Pi_{\rho}(s)$ and $\Pi_{W/\gamma}(s)$
just defined fulfill each a dispersion 
relation~\cite{newpaper} and their imaginary parts are influenced by
SU(3) flavor symmetry breaking. Each of these carries a subtraction
polynomial, which has been chosen of degree 2 and vanishing at the origin.
Additionally, it has been possible to relate the subtraction
polynomials for $\Pi_{W}(s)$ and $\Pi_{\gamma}(s)$. All this lessens
significantly the model parameter freedom.

Breaking isospin symmetry turns out to multiply $|F_\pi^\tau(s)|^2$
by some specific factors~\cite{Marciano,Cirigliano,Mexico} not discussed 
here (see~\cite{newpaper}) and add the $\omg$ and $\phi$ contributions
to  $F_\pi^e(s)$. This, however,  has been shown  
insufficient in order to restore consistency between $e^+e^-$ and $\tau$
data~\cite{DavierPrevious,DavierPrevious3,Davier2007}.

\section{A model for $\rho^0$, $\omg$, $\phi$ mixing}
\label{sect4}
\indent \indent As it is clear that the isospin symmetry breaking effects 
listed above have to be taken into account, the question is rather about
a missing piece in the scheme outlined in Section \ref{sect3}. While working
at one loop order, the HLS model provides self--masses already
referred to for the $\rho$ meson propagators. However, it also contains the piece~: 
\be
\displaystyle \frac{ia g}{4 z_A} \left[
(\rho^0_I + \omg_I -\sqrt{2}  z_V \phi_I ) ~K^- \parsym K^+ 
+
(\rho^0_I - \omg_I +\sqrt{2}  z_V \phi_I ) ~K^0 \parsym \overline{K}^0 
\right]
\label{eq3}
\ee
which -- through kaon loops -- generates transitions~\footnote{
The anomalous and the Yang--Mills pieces of the full HLS Lagrangian
contribute also to the mechanism we outline here in a quite analogous 
manner~\cite{newpaper}; these contributions will be skipped from now on.
The constants $z_A$ and $z_V$ in Eq. \ref{eq3}
are SU(3) breaking parameters which have to be determined by fit.} 
among the so--called ideal (bare) 
fields $\rho^0_I$, $\omg_I$ and $\phi_I$ with no counter part affecting
the $\rho^\pm$ field. We have~:

\be
\left \{
\begin{array}{ll}
 \Pi_{\omega \phi}(s)= -g_{\omega K K} g_{\phi K K} [\Pi_\pm(s)+\Pi_0(s)] 
\\[0.2cm]
\Pi_{\rho \omega}(s)= ~~g_{\rho K K} g_{\omega K K} [\Pi_\pm(s)-\Pi_0(s)]
\\[0.2cm]
\Pi_{\rho \phi}(s)= -g_{\rho K K} g_{\phi K K} [\Pi_\pm(s)-\Pi_0(s)]
\end{array}
\label{eq4}
\right .
\ee
as transition amplitudes between the (ideal) $\rho^0_I$, $\omg_I$ and $\phi_I$.
 $\Pi_\pm(s)$ and $\Pi_0(s)$ denote, resp. the charged and neutral kaon loops
amputated of the coupling constants factored out for sake of clarity.
These loops are defined by dispersion integrals over their imaginary parts and contain
subtraction polynomials ($P_\pm(s)$ and $P_0(s)$) real for real $s$, the invariant mass
squared flowing through the vector meson lines. These polynomials are chosen
of degree 2 and vanishing at $s=0$; their coefficients have to be fixed by 
external conditions. If isospin symmetry is conserved one may assume
that $P_\pm(s)=P_0(s)$ and, then, $\Pi_{\rho \omega}(s)$ and $\Pi_{\rho \phi}(s)$
identically vanish; when isospin symmetry is broken this condition
is certainly no longer fulfilled. Therefore, the HLS model which always predicts 
$\omg_I-\phi_I$ transitions (as $ \Pi_{\omega \phi}(s)$ never vanishes identically), 
predicts additionally $\rho^0_I-\omg_I$ and  $\rho^0_I-\phi_I$
transitions when isospin symmetry is broken.

Therefore, in the general case of isospin symmetry breaking, there are
transitions among the ideal vector fields. If one defines the {\it physical} vector
fields as eigenstates of the vector mass matrix, as the amplitudes in Eqs. \ref{eq4}
provide  non--vanishing entries in the vector meson squared mass matrix,
these cannot coincide with their ideal partners. Let us define the vector $V$
and $V_I$ as the vectors constructed with (resp.) the (physical) $\rho^0$, $\omg$ and $\phi$
fields on the one hand, and the (ideal) $\rho^0_I$, $\omg_I$ and $\phi_I$ fields the other hand. Then
the mass eigenstates of the  vector meson squared mass matrix and the ideal fields are
related by $V=R(s)~V_I$ and $V_I=\widetilde{R(s)} ~V$ with~:
\be
R =
\left (
\begin{array}{lll}
 ~~~~~~~~~1 & \displaystyle \frac{\epsilon_1}{\Pi_{\pi \pi}(s)-\epsilon_2} &
\displaystyle - \frac{\mu \epsilon_1 }{
(1-z_V) m^2 + \Pi_{\pi \pi}(s) -\mu^2 \epsilon_2}\\[0.3cm]
\displaystyle - \frac{\epsilon_1}{\Pi_{\pi \pi}(s)-\epsilon_2} &  ~~~~~~~~~1 &
\displaystyle - \frac{\mu \epsilon_2 }{
(1-z_V) m^2 + (1-\mu^2) \epsilon_2}\\[0.3cm]
\displaystyle   \frac{\mu \epsilon_1 }{(1-z_V) m^2 + \Pi_{\pi \pi}(s) -\mu^2 \epsilon_2}
& \displaystyle\frac{\mu \epsilon_2 }{ (1-z_V) m^2 + (1-\mu^2) \epsilon_2} & \hspace{3.cm} 1
\end{array}
\right ) 
\label{eq5}
\ee
where $\epsilon_1=\Pi_{\rho \omega}(s)$ and $\epsilon_2=\Pi_{\rho \phi}(s)$
are functions of $s$, real  below $ s \sim 1 $ GeV$^2$. Indeed,
the loop imaginary parts start at the corresponding two--kaon thresholds.
One neglects terms of second order in $\epsilon_1$ and/or $\epsilon_2$. 
$ \Pi_{\pi \pi}(s)$ is the pion loop representing the bulk of the $\rho$
self--energy and $m^2=ag^2f_\pi^2$ is the unperturbed $\rho$ meson mass squared.

Performing the change to physical fields into the HLS Lagrangian 
generates~\cite{newpaper}  isospin symmetry violating 
couplings of the $\omg$ and $\phi$ fields to $\pi^+ \pi^-$, while leaving the $\rho^0$ coupling
to $\pi^+ \pi^-$ identical to that of its ideal partner at leading
(first) order in the $\epsilon_i$. In contrast, the $\gamma -\rho^0$ transition amplitude
(named $f_\rho$ in Eq. \ref{eq1}) is modified to~\cite{newpaper} $f_\rho+\delta f_\rho(s)$ 
at leading order in the $\epsilon_i$, while the $W-\rho^\pm$ transition is obviously unaffected. We thus get,
using obvious notations~: 

\be
\left \{
\begin{array}{llll}
\displaystyle f_\rho^\tau &=\displaystyle a g f_\pi^2 
\\[0.3cm]
\displaystyle f_\rho^e &=\displaystyle a g f_\pi^2 \left[
1 + \frac{1}{3} \frac{\epsilon_1}{\Pi_{\pi\pi}(s) -\epsilon_2} 
+\frac{1}{3} \frac{\mu^2 \epsilon_1}{(1-z_V) m^2 +\Pi_{\pi\pi}(s) -\mu^2 \epsilon_2 }
\right]
\end{array}
\right .
\label{eq6}
\ee

Therefore, because of one--loop effects, isospin symmetry breaking  introduces
a $s$--dependent difference between the $\gamma-\rho^0$ and $W-\rho^\pm$ transitions;
this is entirely due to the fact that ideal neutral vector fields cease to coincide 
with physical neutral vector fields, when defined as mass matrix eigenstates. Loop effects always
affect the $(\omg,~\phi)$ sector, but the whole $(\rho^0,~\omg,~\phi)$ is affected only when,
additionally, isospin symmetry is broken. Clearly, this effect has not been accounted for 
in previous analyses of the pion form factor in $e^+e^-$ and $\tau$ data.

\section{How to work  out the model~?} 
\label{sect5}
\indent \indent The issue now
is whether the $(\rho^0,~\omg,~\phi)$ mixing we just sketched is able to 
account {\it numerically} for the long standing mismatch between  
$e^+e^-$ and $\tau$ data. From the point of view of data
analysis, the number of parameters (coupling constants, U(3)/SU(3) breaking
parameters, subtraction parameters from dispersion integrals\ldots) in our HLS
based model is too large to hope fixing them reasonably well
using only the $e^+e^-$ and $\tau$ data. Fortunately, there is a way out.

It indeed happens that the radiative decays ($PV\gamma$ and $P \gamma \gamma$),
which are accounted for by the anomalous sector~\cite{FKTUY} of the HLS Lagrangian,
depend on a large part of the parameters involved in our model and can serve
to fix them quite reliably, even by fitting them in isolation~\cite{rad,mixing}. 
If one adds to this data set the leptonic decay information for
the $\omg$ and $\phi$ mesons on the one hand, and two--pion decay information
of the $\phi$ meson~\footnote{The $e^+e^-$ spectrum in the region of the $\phi$
meson is not available as such in the data published by the various Collaborations at 
Novosibirsk.} on the other hand, the minimization program becomes numerically
well defined. This additional data set will be referred to as "decay data".

Therefore, the resolution method we propose is to consider the $e^+e^-$ and $\tau$ data
together with the decay data. One should stress that~\footnote{ Except for a 
parameter $\delta m_V^2$ which may accout for a (possible) mass difference between $\rho^0$
and $\rho^\pm$.}  the form factor $F_\pi^\tau(s)$ is entirely
determined, from a numerical point of view, by the   $e^+e^-$  and decay data
in isolation, since actually all parameters it depends on are already
involved in the decay widths considered or in  $F_\pi^e(s)$. Stated otherwise,
$F_\pi^\tau(s)$ can be {\it predicted} from our model using only  
the   $e^+e^-$  and decay data. We actually consider this last property as the main
test of validity of our approach.

Before closing this Section, one should mention that the HLS model, as currently
known, involves the pseudoscalar meson  ($P$) and only the lowest lying vector 
meson nonet ($V$). This prevents, for the time being, to include in fit procedures
invariant mass region influenced by higher mass vector mesons, such as the $\rho(1450)$
or  $\rho(1700)$. Therefore, the fits to form factors will be restricted to the $s < 1$
GeV$^2$ region, where the corresponding effects should be limited.
\section{Brief analysis of fit results}
\indent \indent Detailed fit information can be found in Ref.~\cite{newpaper}
where they are lengthily presented and discussed. Here, we limit ourselves
to the most relevant. One should also mention that data on the pion form factor
in the close spacelike region~\cite{NA7,fermilab} are included in our fits.
\begin{table}[ph]  %[t] %
\caption{Global fit information with special emphasis on pion form factor
 data subset contributions.
 \label{chi2:exp}}
\vspace{0.4cm}
\begin{center}
\begin{tabular}{|l|c|c|c|}
\hline
~~ & Full Fit & Excluding $\tau$ data & No $\rho^0-\rho^\pm$ mass shift 
\\ \hline
$\chi^2/dof$ & 313.83/331 & 257.73/274 & 321.75/332\\
Probability  & 74.4\%& 75.2\%& 64.7\%
\\ \hline
\\ \hline
All Timelike Data & & & \\
($\chi^2/$points)& 187.15/(209) & 176.70/(209) &192.38/(209)
\\ \hline
$\tau$ ALEPH & & & \\
($\chi^2/$points) & 23.86/(33)  & {\bf 42.27}/(33) & 24.28/(33)
\\ \hline
$\tau$ CLEO & & & \\
($\chi^2/$points) & 26.06/(25)  & {\bf 26.16}/(25) & 28.55/(25)
\\ \hline
\end{tabular}
\end{center}
\end{table}

 \indent \indent Table \ref{chi2:exp} clearly shows that the description of the global
 data set is quite satisfactory. The second data column gives mostly the $\chi^2$ distance
 of the model to the $\tau$ data points {\it left out from the fit procedure}; this clearly
 illustrates that  $F_\pi^\tau(s)$ is indeed numerically derived from the HLS model together
 with data independent of the $\tau$ form factor. One can even remark that CLEO data~\cite{Cleo}
  are as well accounted for as when including them in the fit procedure~! Comparing the first and 
  third data columns, one may also remark that, allowing a different mass for the charged and neutral
  $\rho$ mesons provides a marginal improvement. Actually, the corresponding mass difference
  (at the edge of statistical significance) is visible only in ALEPH~\cite{Aleph} data and needs
  confirmation by forthcoming data sets.
  
  Therefore, one may conclude that introducing the effects of isospin symmetry breaking
  (a non--zero $\epsilon_1(s)$, substantially) on vector meson mixing,  together with  the already reported
  effects, is enough to reconcile the $e^+e^-$ and $\tau$ data. A missing piece 
  in the current isospin symmetry breaking procedure is then identified as the effects
  of the isospin 0 component of the $\rho^0$ meson which has no counter part inside the 
   $\rho^\pm$ meson. 

\begin{figure}[ph]
\begin{minipage}{0.45\textwidth}
\begin{center}
\resizebox{\textwidth}{!}
{\includegraphics*[width=3cm]{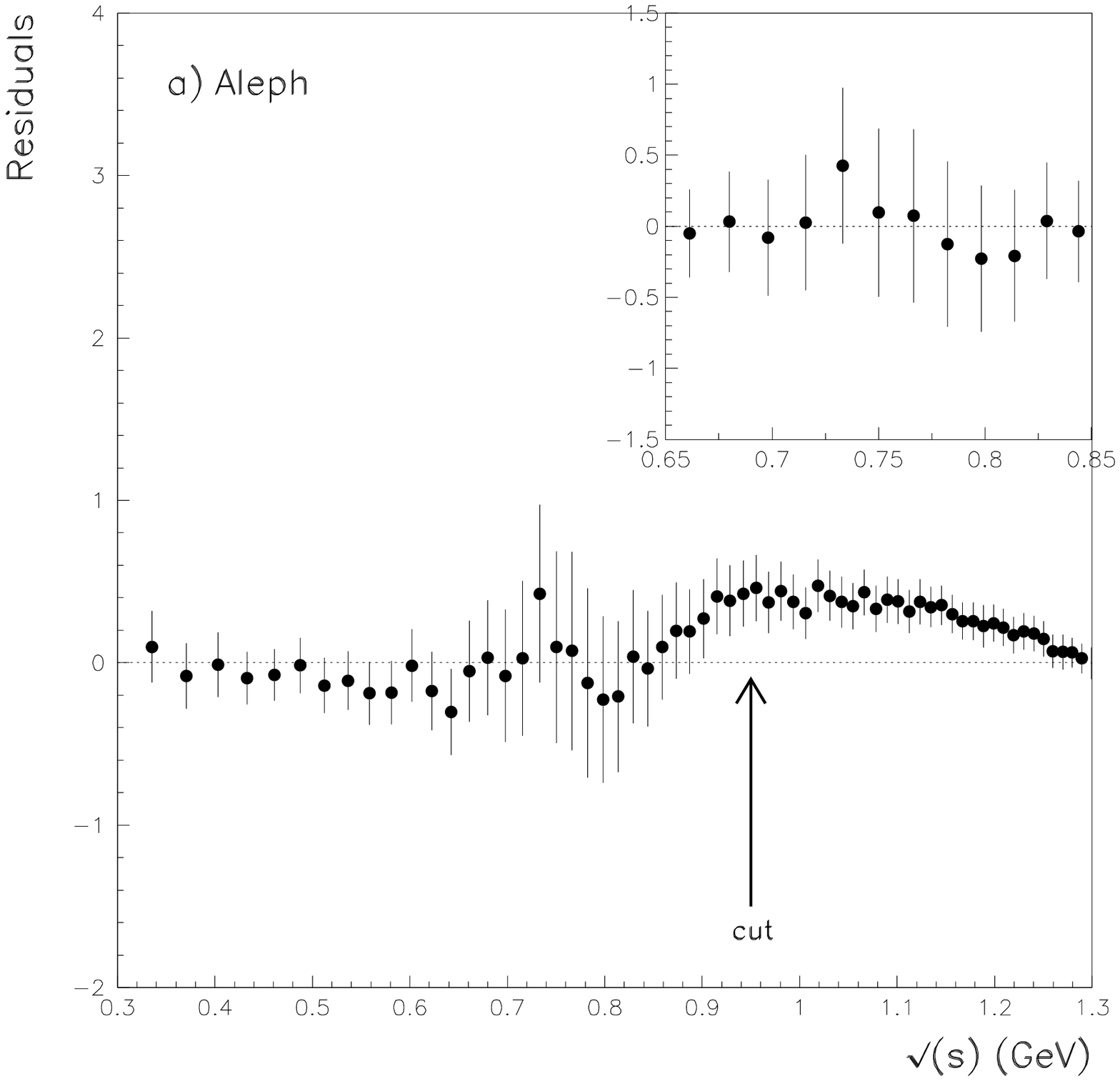}}
\end{center}
\end{minipage}
\begin{minipage}{0.45\textwidth}
\begin{center}
\resizebox{\textwidth}{!}
{\includegraphics*[width=3cm]{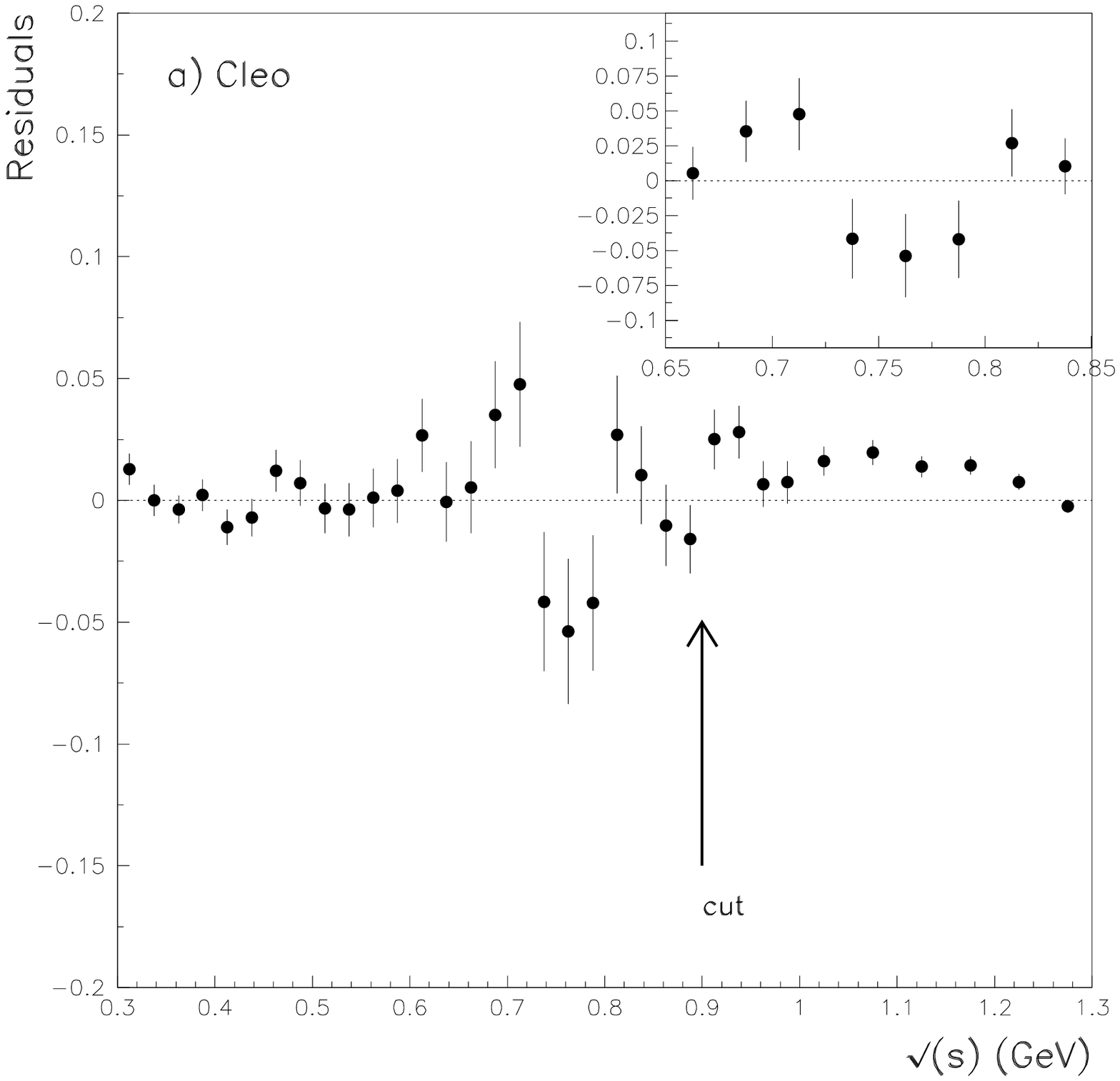}}
%%%{\includegraphics*{light.eps}}
\end{center}
\end{minipage}
\begin{center}
\vspace{-0.5cm}
\caption{\label{TauRes}
Distributions of fit residuals for the  ALEPH and CLEO $\tau$ data
from the full fit with an allowed mass difference between charged
and neutral $\rho$ mesons. The arrows indicate the upper limit
of the fit region.The insets magnify the $\rho$ peak invariant mass region.}
\end{center}
\end{figure}
\begin{figure}[ph]
% \vspace{-1.cm}
					\vspace{-0.9cm}
\begin{center}
\begin{minipage}{0.5 \textwidth}
\resizebox{\textwidth}{!}
{\includegraphics*{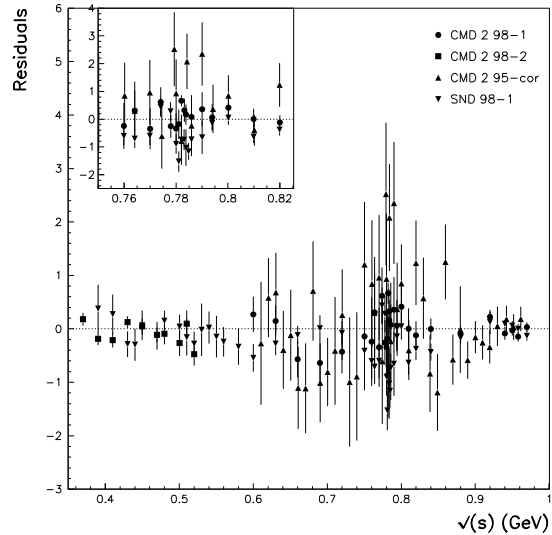}}
\end{minipage}
\end{center}
\vspace{-0.5cm}
\caption{\label{NskRes}
Residual distribution for all the $e^+e^-$ new timelike data over
the whole invariant mass interval. The inset magnifies the $\rho$ peak invariant mass region.}
\end{figure}

\vspace{-0.5cm}
 \indent \indent 
Other information is provided by Figs. \ref{TauRes} and \ref{NskRes} which exhibit the
fit residuals. One can clearly consider them as structureless in the region below 0.9
GeV as no obvious $s$--dependent effect~\cite{Davier2007} can be observed.
One also clearly sees the effects of higher mass vector mesons starting as early
as around the GeV region. 

As final conclusion, one may indeed consider that $e^+e^-$ and $\tau$ data 
do not exhibit any mismatch once all consequences of isospin symmetry breaking
are indeed considered, including the isospin 0 component generated inside
the $\rho^0$ meson. Then, it follows from this work that the
predicted value of the muon anomalous moment derived using $e^+e^-$ data is
indeed reliable and that the actual mismatch is between the prediction of the muon
$g-2$ and its direct (BNL) measurement~\cite{BNL}, rather than between $e^+e^-$ and $\tau$ data.
Therefore, getting an improved measurement~\cite{BNL2} of the muon anomalous magnetic moment
becomes a key issue, possibly a window on some New Physics.

\end{document}